# A Stability Formula for Plastic-Tipped Bullets

## Part 1: Motivation and Development of New Formula

Michael W. Courtney, Ph.D., U.S. Air Force Academy, 2354 Fairchild Drive, USAF Academy, CO, 80840
Michael_Courtney@alum.mit.edu

Donald G. Miller, Ph.D., 2862 Waverley Way, Livermore, CA 94551

**Abstract:** Part 1 of this paper describes a modification of the original Miller twist rule for computing gyroscopic bullet stability that is better suited to plastic-tipped bullets. The original Miller twist rule assumes a bullet of constant density, but it also works well for conventional copper (or gilding metal) jacketed lead bullets because the density of copper and lead are sufficiently close. However, the original Miller twist rule significantly underestimates the gyroscopic stability of plastic-tipped bullets, because the density of plastic is much lower than the density of copper and lead. Here, a new amended formula is developed for the gyroscopic stability of plastic-tipped bullets by substituting the length of just the metal portion for the total length in the $(1 + L^2)$ term of the original Miller twist rule. Part 2 describes experimental testing of this new formula on three plastic-tipped bullets. The new formula is relatively accurate for plastic-tipped bullets whose metal portion has nearly uniform density, but underestimates the gyroscopic stability of bullets whose core is significantly less dense than the jacket.

## Motivation and Background

Just as a football must spin in a spiral to fly point forward rather than tumbling end-over-end, a bullet in flight requires a minimum spin rate to prevent tumbling. This spin is imparted by the barrel's rifling, and the spin rate required for gyroscopic stability depends on a number of complex factors including the aerodynamic overturning moment and the bullet's moments of inertia about the axes parallel and perpendicular to its symmetry axis. A bullet's stability in flight is analogous to a spinning top. Just as a top that spins too slowly or not at all will fall over due to the force of gravity, a bullet that spins too slowly or not at all will tumble end over end due to the force of air drag.

Conservation of angular momentum gives a bullet a certain amount of gyroscopic stability, which makes its axis resistant to the overturning aerodynamic torque. Just as a bicycle is easier to balance when the wheels are spinning faster, a bullet spinning faster is harder to upset. A numerical measure of the bullet's stability, the stability factor $S_g$, is the ratio of the rigidity of the axis of rotation to the magnitude of the overturning aerodynamic torque.(Litz 2009a) More simply stated, the gyroscopic stability factor is the ratio of the tendency to remain point forward to the tendency to tumble in flight. In a perfect world, any stability factor greater than 1.0 would ensure stable flight. Owing to imperfections in bullet construction, barrel manufacturing, knowledge of atmospheric conditions, and uncertainty of formulas, experts recommend selecting a twist rate that provides $S_g$ in the range 1.4-2.0 to ensure stable flight. It has also been observed (Litz 2009a) that a bullet with gyroscopic stability less than 1.3 can exhibit a significant increase in aerodynamic drag.

The 1879 Greenhill formula was widely used as a simple empirical model for predicting bullet stability until the Miller twist rule was published in 2005 (Miller 2005). More complex options for predicting bullet stability include Robert McCoy's McGyro, a program available at http://www.jbmballistics.com/ballistics/downloads/downloads.shtml, and high-end solid-modeling software such as PRODAS (www.prodas.com). These involve bullet dimensions not generally available.

The Miller twist rule has become popular, because it is easy to implement (Litz 2009a) and provides accurate results for many bullets (Miller 2009) from information that is easily available to most shooters. Three useful implementations are at the JBM web site (http://www.jbmballistics.com/cgi-bin/jbmstab-5.1.cgi), at the Accurate Shooter web site http://accurateshooter.net/Blog/millerformula.xls), and

# A Stability Formula for Plastic-Tipped Bullets

a more elaborate Excel one available at don.miller9@comcast.net. The formula is also implemented in Litz's "Point Mass Ballistics Solver" included with his book "Applied Ballistics for Long Range Shooting." (Litz 2009a, 2011).

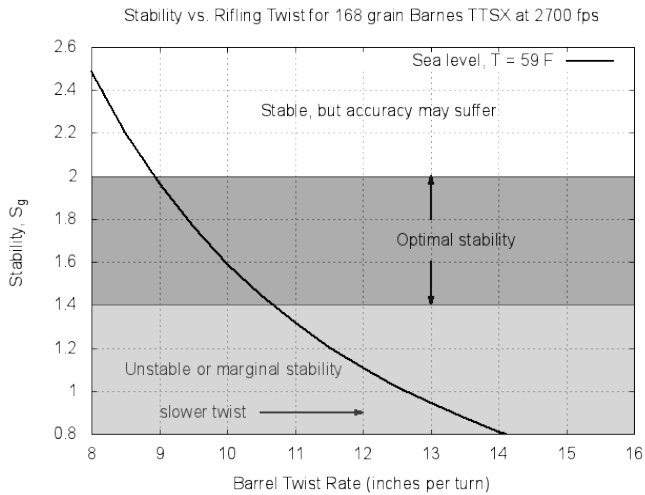

*Figure 1: Predictions of the Miller twist rule for gyroscopic bullet stability vs. rifle twist rate for the Barnes 168 grain tipped TSX. Under standard atmospheric conditions, this bullet is predicted to be optimally stable at twist rates between 9 and 10.5 inches per turn. However, for a 1 in 12" twist rate, this bullet is predicted to be barely stable with a Sg = 1.1. Even if point-forward flight is maintained, this low stability suggests a significant decrease in ballistic coefficient.*

The required inputs for the original Miller twist formula are the bullet weight, bullet length, caliber, muzzle velocity, barrel twist rate, along with atmospheric pressure and ambient temperature, all of which are easily available to the shooter. Except for length, all are already commonly used by ballistic calculators in computing bullet trajectories.

Even though the Miller twist rule has been shown to be accurate for a number of bullets (Miller 2009), it was not expected to be accurate for plastic-tipped bullets, and in fact predicts instability for many of them at commonly used velocities and twist rates where they are known to be stable. This was well known to author Don Miller (DM) and various bullet manufacturers. However, it was somewhat unexpected to author Michael Courtney (MC) when he made a graph analogous to Figure 1 as part of evaluating the 0.308" diameter 168 grain Barnes TTSX for potential use as an elk hunting bullet in his 1 turn in 10" twist 30-06 and 1 turn in 12" twist .308 Winchester.

Not wanting to select a bullet and rifle combination for hunting that might be marginal, MC contacted Barnes inquiring into the gyroscopic stability of this bullet in a 1 in 12" twist, and Barnes copied DM on the reply. A consensus quickly developed that this bullet is indeed expected to be well-stabilized by a 1 in 12" twist rate at .308 Winchester muzzle velocities, and that the failure of the Miller twist rule in this case is because of the bullet's plastic tip.

A productive email exchange then ensued between the authors (DM and MC), where we discussed potential modifications of the twist rule to improve its application to plastic-tipped bullets, as well as the kind of experimental data and range observations that would be needed to gain confidence in a modified formula.

## Original Miller Twist Rule and Modification for Plastic-Tipped Bullets

Because a bullet's moments of inertia are not generally known or easily obtainable, a number of formulas have been offered in attempts to reliably estimate gyroscopic bullet stability (or required rifling twist rate) from easily-available bullet parameters such as weight, length, and muzzle velocity. The Greenhill formula was widely used for many decades, but the Miller formula for bullet stability has been shown to be more accurate and widely applicable for supersonic flight. (Miller 2005, Miller 2009)

The Miller formula for gyroscopic bullet stability is (Equation 1):

$$S_g = \frac{30\,m}{t^2 d^3 L(1+L^2)} \times \left(\frac{V}{2800}\right)^{\frac{1}{3}} \times \frac{(FT+460)}{(59+460)} \frac{29.92}{P},$$

where $m$ is the mass of the bullet in grains, $t$ the twist of the barrel in calibers per turn, $d$ the diameter (caliber) of the bullet in inches, $L$ the length of the bullet in calibers, $V$ is muzzle velocity of the bullet in feet per second, $FT$ is ambient temperature in degrees Fahrenheit, and $P$ is air pressure in inches of mercury.

The first factor in this equation,

$$\frac{30\,m}{t^2 d^3 L(1+L^2)},$$

# A Stability Formula for Plastic-Tipped Bullets

is simply the "uncorrected stability factor," i.e., the stability factor at a muzzle velocity of 2800 fps and a standard atmosphere of 59 °F and 29.92" of mercury. It applies to solid and solid-core bullets (Miller 2005). This part of the formula shows that other factors being equal, stability decreases with increased bullet length, increases with mass, and increases with a faster twist rate (smaller t, fewer calibers per turn).

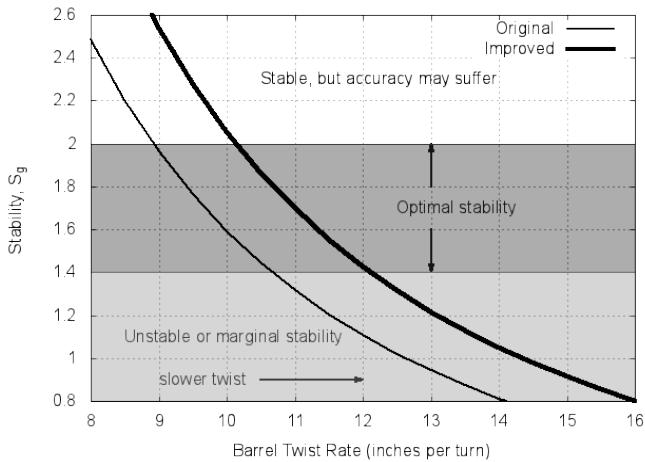

*Figure 2: The improved gyroscopic stability formula (Equation 2) gives results that agree with Barnes' claim that the bullet is well stabilized from a 1 in 12" twist barrel.*

The second factor in the equation,

$$\left(\frac{V}{2800}\right)^{\frac{1}{3}},$$

is the velocity correction factor, describing how stability factor changes for muzzle velocities other than 2800 fps. This correction factor for all velocities *below* the velocity of sound has the value *at* the velocity of sound.

The last factor,

$$\frac{(FT+460)}{(59+460)}\frac{29.92}{P},$$

is the air-density correction factor. It incorporates effects for variations in the ambient temperature and pressure from standard conditions. The most significant effects come from variations in atmospheric pressure with altitude. For example, the atmospheric pressure is P = 24.90" Hg at 5000 ft, and 20.58" Hg at 10,000 ft. (Litz 2009a pp. 533-539)

After considering several possible modifications to the original formula to better estimate the gyroscopic stability of plastic-tipped bullets, we settled on the formula (Equation 2):

$$S_g = \frac{30\,m}{t^2 d^3 L(1+L_m^2)} \times \left(\frac{V}{2800}\right)^{\frac{1}{3}} \times \frac{(FT+460)}{(59+460)}\frac{29.92}{P},$$

where $L_m$ is the length of the *metal portion* of the bullet. This adjustment was most reasonable, because the length in the $(1 + L^2)$ term is related to the rotational moments of inertia. That term will be more accurately estimated by the length of the *metal* part of the bullet because the density of copper and lead are close to 10 times the density of the plastic tip. Consequently, the length added by the plastic tip does little to change the moments of inertia. In contrast, the other $L$ in the formula originates in the center of pressure, an aerodynamic property determined by the shape of the bullet without regard to material density, with a smaller effect from the center of gravity. Thus it makes most sense for this to remain as the total length of the bullet.

**Possible Experiments**

Not satisfied with a formula that merely agrees with a bullet company's claims regarding the gyroscopic stability of a single plastic-tipped bullet design, we considered possible experimental designs to provide a more quantitative test than whether or not bullets tumble and whether predictions agree with claims of bullet manufacturers or anecdotal range reports.

One important consideration came from observations by William McDonald and Ted Almgren (Sierra Loading Manual, 5th ed.), and more specifically by Bryan Litz (Litz 2009a), that a bullet with a gyroscopic stability factor less than 1.3 will demonstrate a reduction in ballistic coefficient (that is, an increase in aerodynamic drag.) Consequently, our experimental design leaned toward measuring the BC of bullets using two chronographs separated by 300 feet, and inferring the decrease in stability factor from 1.3 down to 1.1 from a decrease in ballistic coefficient even before tumbling (key holes in the target) could be directly observed. We also wanted a test method that would be accessible to other shooters, executable with rifles we own or could easily access, and would test each bullet design in a *single* rifle barrel so as not to introduce confounding effects of different rifle barrels on aerodynamic drag. (Courtney and Courtney 2009, Litz 2009b)

# A Stability Formula for Plastic-Tipped Bullets

Our strategy was based on the idea that as gyroscopic stability factors are gradually reduced from 1.3 to 1.1, the observed BC of the bullet should decrease, and then the bullet should tumble as the stability factor nears or crosses 1.0. Since different rifle barrels can also give rise to different BC, using different rifle barrels to observe effects of different twist rates seemed ill-advised, and also impractical since we did not own a number of rifles with different twist rates all in the same chambering. The parameter best suited to adjust stability appeared to be velocity because we could make observations at more stabilities than possible with different rifle barrels.

One way to achieve this goal was to pick several bullets with gyroscopic stability factors close to 1.4 with full-power .223 Remington loads in a 1 in 12" twist barrel, then reduce the charge gradually. A .223 is particularly convenient because one of the authors (MC) owns a Rem 700 he's used for years, and it is commonly known that Blue Dot is usable with a wide range of charges to reduce the muzzle velocity down near 1100 feet per second. The only challenge to this plan was that MC lives near Colorado Springs at an elevation above 7000 ft. The stability of bullets is much higher at that elevation, due to the thinness of the air. A trip to a lower elevation would be necessary.

Part 2 will describe experimental testing of the new gyroscopic stability formula in three plastic-tipped bullets.

**Bibliography**

**Courtney, Michael and Courtney, Amy.** In*accurate Specifications of Ballistic Coefficients.* The Varmint Hunter Magazine, Issue 69. 2009.

**Litz, Brian.** *Applied Ballistics for Long Range Shooting.* Cedar Springs, MI : Applied Ballistics, LLC, 2009a, 2nd Edition, 2011.

**Litz, Brian.** *Accurate Specifications of Ballistic Coefficients.* The Varmint Hunter Magazine, Issue 72. 2009b.

**McDonald, William and Algren, Ted.** *Sierra Loading Manual, 5th ed.*, Section 2.5.

**Miller, Don.** A New Rule for Estimating Rifling Twist: An Aid to Choosing Bullets and Rifles. *Precision Shooting.* March 2005, pp. 43-48.

**Miller, Don.** How Good Are Simple Rules for Estimating Rifle Twist. *Precision Shooting.* June 2009, pp. 48-52.

*Note:* The addresses and affiliations listed are where the work was completed in 2010-2012. Michael Courtney's current affiliation is BTG Research, Baton Rouge, Louisiana.